\documentclass[aps,prd,onecolumn,groupedaddress,showpacs,nofootinbib,amssymb]{revtex4}
\usepackage[dvips]{graphicx}
\usepackage{amssymb}
\usepackage{amsmath}
\usepackage{graphicx,,color}
\usepackage{amsfonts}
\usepackage{bm}
\usepackage{cancel}
\usepackage{comment}

\newcommand\be{\begin{equation}}
\newcommand\ee{\end{equation}}

\newcommand{\ba}{\begin{eqnarray}}
\newcommand{\ea}{\end{eqnarray}}
\allowdisplaybreaks[4]

\allowdisplaybreaks[4]

\begin{document}

\title{Power-law $F(R)$ Gravity as Deformations to Starobinsky Inflation in
View of ACT}
\author{S.D. Odintsov$^{1,2,3,4}$}\email{odintsov@ice.csic.es}
\author{V.K. Oikonomou,$^{4,5}$}\email{voikonomou@gapps.auth.gr;v.k.oikonomou1979@gmail.com}
\affiliation{$^{1)}$ Institute of Space Sciences (ICE, CSIC) C. Can Magrans s/n, 08193 Barcelona, Spain \\
$^{2)}$ ICREA, Passeig Luis Companys, 23, 08010 Barcelona, Spain\\
$^{3)}$Institut d'Estudis Espacials de Catalunya (IEEC), 08860
Castelldefels, Barcelona, Catalonia, Spain\\
$^{4)}$L.N. Gumilyov Eurasian National University - Astana,
010008, Kazakhstan \\
$^{5)}$Department of Physics, Aristotle University of
Thessaloniki, Thessaloniki 54124, Greece}

\tolerance=5000

\begin{abstract}
In this work we aim to provide a new parametrization of power-law
$F(R)$ gravity inflation framework in the Jordan frame. It is
known in the literature that the power-law $F(R)$ gravity
inflation of the form $F(R)=R+\beta R^n$ is non-viable and
produces a power-law evolution. We demonstrate that the standard
approach in power-law $F(R)$ gravity inflation has some
parametrization issues that may lead to inconsistencies and we
introduce a new parametrization which elevates the role of
power-law $F(R)$ gravity deformations of the Starobinsky
inflation, making it viable and compatible with both the Planck
and ACT data. In our approach the power-law $F(R)$ gravity
inflation is disentangled from a power-law evolution.
\end{abstract}


\maketitle

\section{Introduction}

General relativity (GR) has passed numerous observational tests
over the years, however cracks in its successful description of
the late-time era have started to appear for some years now. The
$\Lambda$-Cold-Dark-Matter model is the benchmark model for the
late Universe, if one sticks in the framework of GR. However, the
latest DESI data \cite{DESI:2024uvr} indicated that the dark
energy is dynamical and also a phantom crossing occurs
\cite{Lee:2025pzo,Ozulker:2025ehg,Kessler:2025kju,Nojiri:2025low,Vagnozzi:2019ezj},
not to mention the Hubble tension problems of the
$\Lambda$-Cold-Dark-Matter model
\cite{Pedrotti:2024kpn,Jiang:2024xnu,Vagnozzi:2023nrq,Adil:2023exv,Bernui:2023byc,Gariazzo:2021qtg}.
These two features cannot be consistently described in the context
of simple GR, unless one invokes phantom scalar fields, which are
in plain words not appealing for a physical description of the
Universe. On the other hand, $F(R)$ gravity
\cite{reviews1,reviews2,reviews3,reviews4} serves as the simple
and Occam's razor modification of GR. The way of thinking is
simple, the GR action contains the Ricci scalar, so if one thinks
of a modification of GR, the simplest extension is a function of
the Ricci scalar that also contains the linear Einstein-Hilbert
term. In the literature there appear various works on $F(R)$
gravity, for a mainstream of works in this context see
\cite{Nojiri:2003ft,Capozziello:2005ku,Capozziello:2004vh,Capozziello:2018ddp,Hwang:2001pu,Cognola:2005de,Nojiri:2006gh,Song:2006ej,Capozziello:2008qc,Bean:2006up,Capozziello:2012ie,Faulkner:2006ub,Olmo:2006eh,Sawicki:2007tf,Faraoni:2007yn,Carloni:2007yv,
Nojiri:2007as,Capozziello:2007ms,Deruelle:2007pt,Appleby:2008tv,Dunsby:2010wg,Odintsov:2020nwm,Odintsov:2019mlf,Odintsov:2019evb,Oikonomou:2020oex,Oikonomou:2020qah,Huang:2013hsb,Berry:2011pb,Bonanno:2010bt,Gannouji:2008wt,Oyaizu:2008sr,Oyaizu:2008tb,Brax:2008hh,Cognola:2007zu,Boehmer:2007glt,Boehmer:2007kx,
deSouza:2007zpn,Song:2007da,Brookfield:2006mq,delaCruz-Dombriz:2006kob,Achitouv:2015yha,Kopp:2013lea,Sebastiani:2013eqa,Odintsov:2017hbk,Myrzakulov:2015qaa,Feng:2022vcx}
and references therein.

In the standard literature, power-law $F(R)$ gravity of the form
$F(R)=R+\beta R^n$ is considered a non-viable inflationary model,
in the slow-roll approximation. In this work we aim to discuss why
the standard treatment of power-law $F(R)$ gravity in the
literature is wrong. We highlight the reasons why the standard
approach of power-law $F(R)$ gravity has inconsistent
parametrizations, and we present a new parametrization for $F(R)$
gravity inflation that can consistently describe power-law $F(R)$
gravity inflation. As we show, the reformed parametrization of
power-law $F(R)$ gravity we will develop is compatible with the
Planck data \cite{Planck:2018jri}, the ACT data
\cite{ACT:2025fju,ACT:2025tim} and the updated Planck constraints
on the tensor-to-scalar ratio \cite{BICEP:2021xfz}. Recall that
the ACT data combined with the DESI data \cite{DESI:2024uvr} yield
a scalar spectral index,
\begin{equation}\label{act}
n_{\mathcal{S}}=0.9743 \pm
0.0034,\,\,\,\frac{\mathrm{d}n_{\mathcal{S}}}{\mathrm{d}\ln
k}=0.0062 \pm 0.0052\, .
\end{equation}
and the updated Planck constraint on the tensor-to-scalar ratio
yields \cite{BICEP:2021xfz},
\begin{equation}\label{planck}
r<0.036
\end{equation}
at $95\%$ confidence. The results of ACT already created a large
stream of articles aiming to find models compatible with the ACT
data
\cite{Kallosh:2025rni,Gao:2025onc,Liu:2025qca,Yogesh:2025wak,Yi:2025dms,Peng:2025bws,Yin:2025rrs,Byrnes:2025kit,Wolf:2025ecy,Aoki:2025wld,Gao:2025viy,Zahoor:2025nuq,Ferreira:2025lrd,Mohammadi:2025gbu,Choudhury:2025vso,Odintsov:2025wai,Q:2025ycf,Zhu:2025twm,Kouniatalis:2025orn,Hai:2025wvs,Dioguardi:2025vci,Yuennan:2025kde,Kuralkar:2025zxr,Kuralkar:2025hoz,Modak:2025bjv,Oikonomou:2025xms,Oikonomou:2025htz,Odintsov:2025jky,Aoki:2025ywt},
although the ACT data should be considered with some restraint
\cite{Ferreira:2025lrd}. Since inflation in its various forms
\cite{inflation2,inflation3,inflation4} will soon be further
tested by stage 4 Cosmic Microwave Background experiments like the
Simons observatory \cite{SimonsObservatory:2019qwx}, and also by
future gravitational wave experiments
\cite{Hild:2010id,Baker:2019nia,Smith:2019wny,Crowder:2005nr,Smith:2016jqs,Seto:2001qf,Kawamura:2020pcg,Bull:2018lat,LISACosmologyWorkingGroup:2022jok},
our analysis offers another possibility of a viable inflationary
phenomenology, that of power-law $F(R)$ gravity, which was
considered not viable.

\section{The Parametrization Issues of Standard Power-law $F(R)$ Gravity in the Jordan Frame}

Let us analyze in depth the problem with the standard approach of
power-law $F(R)$ gravity in the Jordan frame. We consider the
vacuum $F(R)$ gravity, with action,
\begin{equation}\label{FRgravitygeneralaction}
\mathcal{S}=\frac{1}{2\kappa^2}\int \mathrm{d}^4x\sqrt{-g}F(R)\, ,
\end{equation}
with $\kappa^2=8\pi G=\frac{1}{M_p^2}$, where $M_p$ is the reduced
Planck mass, and $G$ is Newton's constant. Varying the action with
respect to the metric, we get,
\begin{equation}\label{eqnmotion}
F_R(R)R_{\mu \nu}(g)-\frac{1}{2}F(R)g_{\mu
\nu}-\nabla_{\mu}\nabla_{\nu}F_R(R)+g_{\mu \nu}\square F_R(R)=0\,
,
\end{equation}
with $F_R=\frac{\mathrm{d}F}{\mathrm{d}R}$. Eq. (\ref{eqnmotion})
can be recast as follows,
\begin{align}\label{modifiedeinsteineqns}
R_{\mu \nu}-\frac{1}{2}Rg_{\mu
\nu}=\frac{\kappa^2}{F_R(R)}\Big{(}T_{\mu
\nu}+\frac{1}{\kappa^2}\Big{(}\frac{F(R)-RF_R(R)}{2}g_{\mu
\nu}+\nabla_{\mu}\nabla_{\nu}F_R(R)-g_{\mu \nu}\square
F_R(R)\Big{)}\Big{)}\, .
\end{align}
For a FRW metric,
\begin{equation}
\label{JGRG14} ds^2 = - dt^2 + a(t)^2 \sum_{i=1,2,3}
\left(dx^i\right)^2\, ,
\end{equation}
the field equations take the form,
\begin{align}
\label{JGRG15} 0 =& -\frac{F(R)}{2} + 3\left(H^2 + \dot H\right)
F_R(R) - 18 \left( 4H^2 \dot H + H \ddot H\right) F_{RR}(R)\, ,\\
\label{Cr4b} 0 =& \frac{F(R)}{2} - \left(\dot H +
3H^2\right)F_R(R) + 6 \left( 8H^2 \dot H + 4 {\dot H}^2 + 6 H
\ddot H + \dddot H\right) F_{RR}(R) + 36\left( 4H\dot H + \ddot
H\right)^2 F_{RRR} \, ,
\end{align}
with $F_{RR}=\frac{\mathrm{d}^2F}{\mathrm{d}R^2}$, and in addition
$F_{RRR}=\frac{\mathrm{d}^3F}{\mathrm{d}R^3}$. The Ricci scalar
which for the FRW metric is,
\begin{equation}\label{ricciscalarfrw}
R=12H^2 + 6\dot H\, .
\end{equation}

Now let us consider the standard treatment of power-law $F(R)$
gravity and let us highlight the parametrization issues that may
lead to inconsistencies. We shall assume a slow-roll regime, so
\begin{equation}\label{slowrollconditionshubble}
\ddot{H}\ll H\dot{H},\,\,\, \frac{\dot{H}}{H^2}\ll 1\, ,
\end{equation}
hence primordially, the Ricci scalar is approximately equal to,
\begin{equation}\label{ricciscalarapprox}
R\sim 12 H^2\, ,
\end{equation}
due to the slow-roll assumption $\frac{\dot{H}}{H^2}\ll 1$. Now
let us dwell in the core of the problem at hand, so let us
consider a power-law $F(R)$ gravity, of the form,
\begin{equation}\label{polynomialfr} f(R)=R+\beta R^n\, ,
\end{equation}
for $n$ a real arbitrary number. The Friedmann equation of the
vacuum $F(R)$ gravity takes the form,
\begin{equation}\label{friedmannewappendix}
3 H^2F_R=\frac{RF_R-F}{2}-3H\dot{F}_R\, ,
\end{equation}
where $F_R=\frac{\partial F}{\partial R}$. During the inflationary
era, we have approximately that $F_R\sim n\beta R^{n-1}$ therefore
the Friedmann equation (\ref{friedmannewappendix}) approximately
becomes,
\begin{align}\label{eqnsofmkotionfrpolyappendix}
& 3 H^2n\beta R^{n-1}=\frac{\beta (n-1)R^{n-1}}{2}-3n(n-1)\beta
HR^{n-2}\dot{R}\, .
\end{align}
Now following what is known to be the standard approach in the
$F(R)$ gravity literature, utilizing the slow-roll approximation,
the Ricci scalar $R=12H^2+6\dot{H}$ becomes $R\sim 12 H^2$ during
inflationary slow-roll regime at leading order, hence the
Friedmann equation (\ref{eqnsofmkotionfrpolyappendix}) becomes,
\begin{equation}\label{leadingordereqnappendix}
3H^2n\beta \simeq 6\beta (n-1)H^2-6n\beta(n-1)\dot{H}+3\beta
(n-1)\dot{H}\, .
\end{equation}
Eq. (\ref{leadingordereqnappendix}) can be solved, analytically
with the solution being,
\begin{equation}\label{hubblefrpolyappendix}
H(t)=\frac{1}{p\,t}\, ,
\end{equation}
with $p=\frac{2-n}{(n-1)(2n-1)}$. The solution basically describes
a simple inverse power-law behavior, which can be an inflationary
evolution only if $1.36<n<2$. At this point, let us indicate the
shortcomings of what is considered to be the standard approach in
power-law $F(R)$ gravity inflation. In a nutshell, the problems
are:
\begin{itemize}\label{problems}
 \item The slow-roll approximation is violated
 \item The analysis cannot produce the physics of the Starobinsky
 inflation, which is a power-law $F(R)$ gravity with a quasi-de Sitter evolution.
 \item Small deformations of the Starobinsky inflation are not
 viable. In fact, some of these do not describe inflation.
 \item Power-law $F(R)$ gravity inflation in this formalism is not
 viable.
 \item In this formalism, $\dot{\epsilon}_1=0$, thus inflation is
 eternal at least classically. This is because quantum corrections
 can end inflation even if one has a theory which classically
 yields $\dot{\epsilon}_1=0$. Hence in the standard treatment of
 power-law $F(R)$ gravity, one ends up with eternal inflation
 classically.
\end{itemize}
Let us start with the first problem, and let us recall that the
power-law evolution of Eq. (\ref{hubblefrpolyappendix}) was
obtained by making the assumption $\dot{H}\ll H^2$, during the
slow-roll era.  However for the power-law evolution of Eq.
(\ref{hubblefrpolyappendix})  we get $\dot{H}=-p\,H^2$. Hence if
we set simply $n=1.37$ one obtains $p=0.978565$, which clearly
violates abruptly the slow-roll condition. Therefore, the solution
itself, violates the slow-roll condition. Secondly, the case $n=2$
cannot be produced in this formalism. This is a serious issue,
since the number $n$ is an arbitrary number, so the case $n=2$
should be normally derived in this formalism. It turns out that
the case $n=2$ does not even describe inflation, and in fact in
this formalism, the case $n=2$ is a non-viable power-law
evolution. But even small deformations of the Starobinsky
inflation in this context are not viable, and some of which are
not even inflationary eras. Let us explain these two issues in
some detail to make the argument clearer. In the case that we
consider $R^{2+\epsilon}$ gravity with $n=2+\epsilon$ and
$\epsilon\ll 1$, according to this formalism, this gravity results
to a power-law evolution, which cannot describe inflation (recall
$1.36<n<2$). This is a major issue. Coming to the problem of small
deformations again, one expects that slight deformations of the
Starobinsky inflation of the form $R^{2-\epsilon}$ with
$\epsilon\ll 1$, should in principle be deformations of an
inflationary quasi-de Sitter regime, and of course these should be
viable deformations. Consider for example $n=1.999$, so basically
$\epsilon\ll 1$. The current approach for power-law $F(R)$ gravity
yields the following slow-roll indices,
\begin{equation}\label{epsiloniforfrpoly}
\epsilon_1=\frac{2-n}{(n-1)(2n-1)},\,\,\,\epsilon_3=-(n-1)\epsilon_1,\,\,\,\epsilon_4=\frac{n-2}{n-1}\,
,
\end{equation}
and also the observational indices are in this case,
\begin{equation}\label{vacuumspectral}
n_s=1-4\epsilon_1+2\epsilon_3-2\epsilon_4,\,\,\,r=48
\frac{\epsilon_3^2}{(1+\epsilon_3)^2}\, .
\end{equation}
For $n=1.999$ we get $n_s=0.999999$, which is not reasonable, and
contradicts intuition which states that slight deformations of the
$R^2$ model should be quasi-de Sitter deformations. But this
formalism results to non-viable models in general. For example, if
$n=1.8$ we get $n_s=0.961538$ and $r=0.3333$ and for $n=1.84$ we
get $n_s=0.977257$ and $r=0.1935$, which is excluded by both the
Planck 2018 and ACT data. Hence the source of the problem is that
this formalism cannot produce realistic physical outcomes for
power-law $F(R)$ gravity. Also using the formalism of this
section, one ends up to an inflationary regime with
$\dot{\epsilon}_1=0$, thus inflation is eternal, at least
classically. In the next section, we revisit the power-law $F(R)$
gravity inflation using a more concrete and compatible with
intuition approach, which rectifies all the problems we discussed
in this section.

\section{Revisited Power-law $F(R)$ Gravity in the Jordan Frame and ACT: New Parametrization}

\subsection{General $F(R)$ Gravity Inflation in the Jordan Frame}

In standard $F(R)$ gravity texts, the inflationary dynamical
evolution is mainly quantified in terms of the slow-roll indices,
$\epsilon_1$ ,$\epsilon_2$, $\epsilon_3$, $\epsilon_4$. The
analytic form of these slow-roll indices for $F(R)$ gravity are
\cite{Hwang:2005hb,reviews1,Odintsov:2020thl,Oikonomou:2025qub},
\begin{equation}
\label{restofparametersfr}\epsilon_1=-\frac{\dot{H}}{H^2}\, ,
\quad \epsilon_3= \frac{\dot{F}_R}{2HF_R}\, ,\quad
\epsilon_4=\frac{\ddot{F}_R}{H\dot{F}_R}\,
 .
\end{equation}
Assuming that the slow-roll indices are  $\epsilon_i\ll 1$,
$i=1,3,4$, the spectral index of the primordial scalar curvature
perturbations and the tensor-to-scalar ratio for $F(R)$ gravity
are \cite{reviews1,Hwang:2005hb},
\begin{equation}
\label{epsilonall} n_s=
1-4\epsilon_1+2\epsilon_3-2\epsilon_4,\quad
r=48\frac{\epsilon_3^2}{(1+\epsilon_3)^2}\, .
\end{equation}
Note that the expression for the tensor-to-scalar ratio easily
follows, if we  consider the ratio of the tensor over scalar power
spectrum,
\begin{equation}\label{tensorananalytic}
r=\frac{P_T}{P_S}=8 \kappa^2 \frac{Q_s}{F_R}\, ,
\end{equation}
with,
\begin{equation}
\label{qsfrpreliminary}
Q_s=\frac{3\dot{F_R}^2}{2F_RH^2\kappa^2(1+\epsilon_3)^2}\, .
\end{equation}
Using Eqs. (\ref{tensorananalytic}) and (\ref{qsfrpreliminary}) we
obtain,
\begin{equation}\label{ranalyticfinal}
r=48 \frac{\dot{F_R}^2}{4F_R^2H^2(1+\epsilon_3)^2}\, ,
\end{equation}
and since $\epsilon_3= \frac{\dot{F}_R}{2HF_R}$, we finally
obtain,
\begin{equation}\label{ranalyticfinal1}
r=48\frac{\epsilon_3^2}{(1+\epsilon_3)^2}\, .
\end{equation}
Directly from the Raychaudhuri equation for a vacuum $F(R)$
gravity, we get the exact equation,
\begin{equation}\label{approx1}
\epsilon_1=-\epsilon_3(1-\epsilon_4)\, .
\end{equation}
This is a very useful equation which we shall utilize in the end
of this section. At leading order we have $\epsilon_1\simeq
-\epsilon_3$, an approximation which will not change our analysis
drastically, since it is a leading order result, so in the
slow-roll regime does not affect our findings. Therefore, in view
of $\epsilon_1\simeq -\epsilon_3$, the spectral index takes the
form,
\begin{equation}
\label{spectralfinal} n_s\simeq 1-6\epsilon_1-2\epsilon_4\, ,
\end{equation}
and also the tensor-to-scalar ratio takes the form,
\begin{equation}
\label{tensorfinal} r\simeq 48\epsilon_1^2\, .
\end{equation}
Now, the detailed calculation of the fourth slow-roll index,
namely $\epsilon_4$, is very important for our analysis. Let us
calculate it in detail, so we have,
\begin{equation}\label{epsilon41}
\epsilon_4=\frac{\ddot{F}_R}{H\dot{F}_R}=\frac{\frac{d}{d
t}\left(F_{RR}\dot{R}\right)}{HF_{RR}\dot{R}}=\frac{F_{RRR}\dot{R}^2+F_{RR}\frac{d
(\dot{R})}{d t}}{HF_{RR}\dot{R}}\, ,
\end{equation}
but note that $\dot{R}$ is,
\begin{equation}\label{rdot}
\dot{R}=24\dot{H}H+6\ddot{H}\simeq 24H\dot{H}=-24H^3\epsilon_1\, ,
\end{equation}
and we took into account the slow-roll approximation $\ddot{H}\ll
H \dot{H}$. Using Eqs. (\ref{rdot}) and (\ref{epsilon41}), we
obtain,
\begin{equation}\label{epsilon4final}
\epsilon_4\simeq -\frac{24
F_{RRR}H^2}{F_{RR}}\epsilon_1-3\epsilon_1+\frac{\dot{\epsilon}_1}{H\epsilon_1}\,
,
\end{equation}
but note that $\dot{\epsilon}_1$ is equal to,
\begin{equation}\label{epsilon1newfiles}
\dot{\epsilon}_1=-\frac{\ddot{H}H^2-2\dot{H}^2H}{H^4}=-\frac{\ddot{H}}{H^2}+\frac{2\dot{H}^2}{H^3}\simeq
2H \epsilon_1^2\, ,
\end{equation}
therefore at leading order, $\epsilon_4$ becomes,
\begin{equation}\label{finalapproxepsilon4}
\epsilon_4\simeq -\frac{24
F_{RRR}H^2}{F_{RR}}\epsilon_1-\epsilon_1\, .
\end{equation}
Thus $\epsilon_4$ may be expressed in terms of the parameter $x$,
which is dimensionless, defined as follows,
\begin{equation}\label{parameterx}
x=\frac{48 F_{RRR}H^2}{F_{RR}}\, .
\end{equation}
Thus in terms of $x$, $\epsilon_4$ takes the form,
\begin{equation}\label{epsilon4finalnew}
\epsilon_4\simeq -\frac{x}{2}\epsilon_1-\epsilon_1\, .
\end{equation}
It is worth at this point, briefly discussing the leading order
approximations that were made. The approximation $\epsilon_1\simeq
-\epsilon_3$ below Eq. (\ref{approx1}) holds true at leading
order, and to be specific at first order in the slow-roll
perturbative expansion of the index $\epsilon_1$ expressed in
terms of $\epsilon_3$ and $\epsilon_4$, so we omitted the term
$\sim \epsilon_3\epsilon_4$. Also it must be noted that no
truncation in terms of the parameter $x$ was made in order to
obtain Eq. (\ref{epsilon4finalnew}). The parameter $x$ arises from
Eq. (\ref{epsilon4final}), if we omit terms which contain the
second derivative of the Hubble rate $\ddot{H}$.

Now using Eqs. (\ref{epsilon4finalnew}) and (\ref{spectralfinal}),
the spectral index of the primordial scalar perturbations takes
becomes,
\begin{equation}\label{asxeto1}
n_s-1=-4\epsilon_1+x\epsilon_1\, .
\end{equation}
and by solving we get,
\begin{equation}\label{spectralasfunctionofepsilon1}
\epsilon_1=\frac{1-n_s}{4-x}\, ,
\end{equation}
so substituting $\epsilon_1$ in the expression for the
tensor-to-scalar ratio in Eq. (\ref{tensorfinal}), we get,
\begin{equation}\label{mainequation}
r\simeq \frac{48 (1-n_s)^2}{(4-x)^2}\, .
\end{equation}
Also let us note that, the parameter $x$ defined in
(\ref{parameterx}) can be expressed in terms of $R$, by
remembering that during the slow-roll inflationary regime we have
$R\sim 12 H^2$, thus,
\begin{equation}\label{parameterxfinal}
x=\frac{4 F_{RRR}\,R}{F_{RR}}\, .
\end{equation}
Hence in this formalism, it is necessary to calculate $x$ and
$\epsilon_1$ at first horizon crossing, and the inflationary
phenomenology is easily evaluated by using Eqs.
(\ref{mainequation}) and (\ref{asxeto1}).

Also note that for a general $F(R)$ gravity, the viability
criteria are,
\begin{equation}\label{criterion1}
F_R>0
\end{equation}
which avoids anti-gravity, and in addition,
\begin{equation}\label{criterion2}
F_{RR}>0
\end{equation}
required from local solar system tests. In addition, if we require
the existence of a stable de Sitter solution during the slow-roll
era, we must require,
\begin{equation}\label{criterion3}
0< y \leq 1\, ,
\end{equation}
with $y$ being,
\begin{equation}\label{yparameterdefinition}
y=\frac{R\,F_{RR}}{F_R}\, .
\end{equation}
The de Sitter existence criterion discussed above, is derived by
simply  perturbing the field equations for the FRW spacetime. If
$R=R_0+G(R)$ is the perturbation, with $R_0$ being the de Sitter
scalar curvature, the Einstein frame scalaron field obeys,
\begin{equation}\label{scalaronequation}
\square G+m^2 G=0\, ,
\end{equation}
where the scalaron mass is \cite{Muller:1987hp},
\begin{equation}\label{scalaronmassinitial}
m^2=\frac{1}{3}\left(-R+\frac{F_R}{F_{RR}} \right)\, .
\end{equation}
The scalaron mass can be expressed in terms of  $y$,
\begin{equation}\label{scalaronmassfinal}
m^2=\frac{R}{3}\left(-1+\frac{1}{y} \right)\, .
\end{equation}
Hence, in order for the stability of the de Sitter point to be
ensured, the scalaron mass must be positive, thus
\begin{equation}\label{desitterscalaron}
0< y \leq 1\, .
\end{equation}

\subsection{The Parametrization of Power-law $F(R)$ Gravity Inflation in the Jordan Frame}

In this subsection we shall use the formalism of the previous
section to present the rectified formalism for studying power-law
$F(R)$ gravity inflation. To start with, if the parameter $x$
defined in Eq. (\ref{parameterxfinal}) is constant, say $x=n$, the
differential equation (\ref{parameterxfinal}) can be solved
analytically and the solution is,
\begin{equation}\label{constnatxol}
F(R)=c_3 R+c_2+\frac{16 c_1 R^{2+\frac{n}{4}}}{(n+4) (n+8)}\, ,
\end{equation}
and this is exactly a power-law $F(R)$ gravity evolution, with
$c_1$, $c_2$ and $c_3$ being appropriate dimensionful integration
constants. Note that in order for the criteria (\ref{criterion1})
and (\ref{criterion2}) to be satisfied, for the scenario of
Starobinsky deformations, one must have $c_1>0$. Indeed
$F'(R)=\frac{4 c_1 R^{\frac{n}{4}+1}}{n+4}+c_3$ and
$F''(R)=\frac{16 c_1 \left(\frac{n}{4}+1\right)
\left(\frac{n}{4}+2\right) R^{n/4}}{(n+4) (n+8)}$, thus $c_1>0$ is
required.

As we showed in the previous subsection, the physics of the
power-law $F(R)$ gravity must be disentangled completely from a
pure power-law evolution of the form $H\sim 1/(p\,t)$. Let us
start from relation (\ref{approx1}), which is an exact relation
extracted by the field equations. We quote it here too for
convenience, hence the starting point is the following equation,
\begin{equation}\label{start}
\epsilon_1=-\epsilon_3(1-\epsilon_4)\, .
\end{equation}
Now, recall that the slow-roll parameter $\epsilon_4$ is at
leading order $\mathcal{O}(\epsilon_1)$ given in Eq.
(\ref{epsilon4finalnew}), thus $\epsilon_4\simeq
-\frac{x}{2}\epsilon_1-\epsilon_1$. Also, from Eq.
(\ref{restofparametersfr}) we can further express the parameter
$\epsilon_3$ in terms of $\epsilon_1$. Indeed we have,
\begin{equation}\label{epsilon3new}
\epsilon_3= \frac{\dot{F}_R}{2HF_R}=\frac{F_{RR}\dot{R}}{2HF_R}\,
,
\end{equation}
and due to the fact that during inflation we have $R\sim 12H^2$,
we have $\dot{R}\sim 24 H\dot{H}$, thus Eq. (\ref{epsilon3new})
yields,
\begin{equation}\label{epsilon3final}
\epsilon_3=-\frac{F_{RR}R}{F_R}\epsilon_1=-y\epsilon_1\, ,
\end{equation}
where we also used Eq. (\ref{yparameterdefinition}). Thus by
combining Eqs. (\ref{start}), (\ref{epsilon3final}) and
(\ref{epsilon4finalnew}), we obtain the following relation,
\begin{equation}\label{epsilon1final1}
\epsilon_1=\frac{2(1-y)}{y(n+2)}\, ,
\end{equation}
and recall that $x=n$ is a constant. Now apparently, the evolution
is not a power-law one, since the parameter $y$ for the $F(R)$
gravity of Eq. (\ref{constnatxol}) is,
\begin{equation}\label{yform}
y=\frac{(4+n)c_1\,R^{1+\frac{n}{4}}}{c_1\,R^{1+\frac{n}{4}}+(4+n)c_3}
\end{equation}
thus for sure we have $\dot{\epsilon}_1\neq 0$ in this case.
Working out and simplifying the functional form of the parameter
$\epsilon_1$ in Eq. (\ref{epsilon1final1}), we get,
\begin{equation}\label{epsilon1final2}
\epsilon_1=-\frac{2 n}{n^2+6 n+8}+\frac{2 c_3
R^{-\frac{n}{4}-1}}{c_1 (n+2)}\, .
\end{equation}
So by assuming that $|\frac{n}{4}|<1$, we can have an estimate for
the value of the parameter $\epsilon_1$ at leading order, and it
is equal to $\epsilon_1\simeq -\frac{2 n}{n^2+6 n+8}$. Note that
this is just a leading order value since the term $\sim
R^{-\frac{n}{4}-1}$ is subdominant during inflation. But still, we
have $\dot{\epsilon}_1\neq 0$, and the leading order value of
$\epsilon_1$ is $\epsilon_1\simeq -\frac{2 n}{n^2+6 n+8}$. Having
the leading order value of $\epsilon_1$ during inflation, we can
proceed to examining the phenomenology of this model, using Eqs.
(\ref{asxeto1}) and (\ref{mainequation}).

\subsubsection{Confrontation with the ACT Data}

\begin{figure}
\centering
\includegraphics[width=28pc]{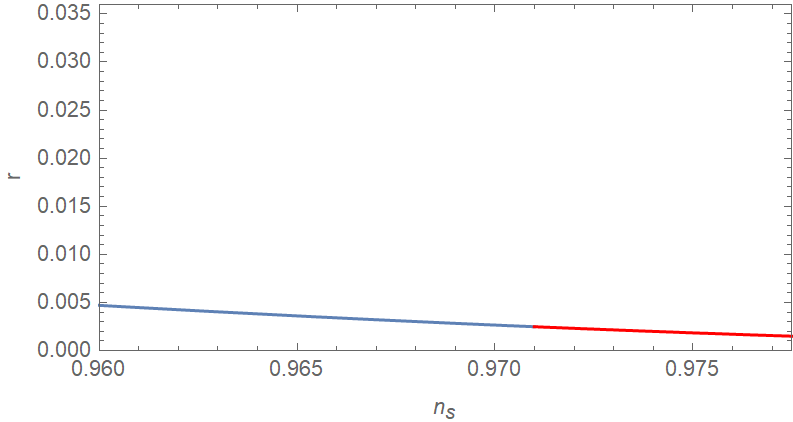}
\caption{Parametric plot of the spectral index and the
tensor-to-scalar ratio for $n$ in the range
$n=[-0.038,-0.022]$.}\label{plot1}
\end{figure}

\begin{figure}
\centering
\includegraphics[width=28pc]{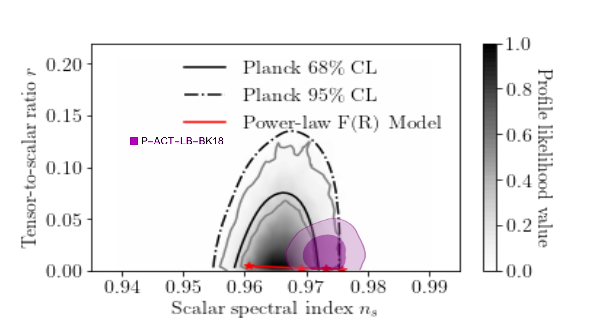}
\caption{Marginalized curves of the Planck 2018 data and the
power-law $F(R)$ gravity model, confronted with the ACT data, the
Planck 2018 data, and the updated Planck constraints on the
tensor-to-scalar ratio for $n=[-0.038,-0.022]$. }\label{plot2}
\end{figure}

\begin{figure}
\centering
\includegraphics[width=20pc]{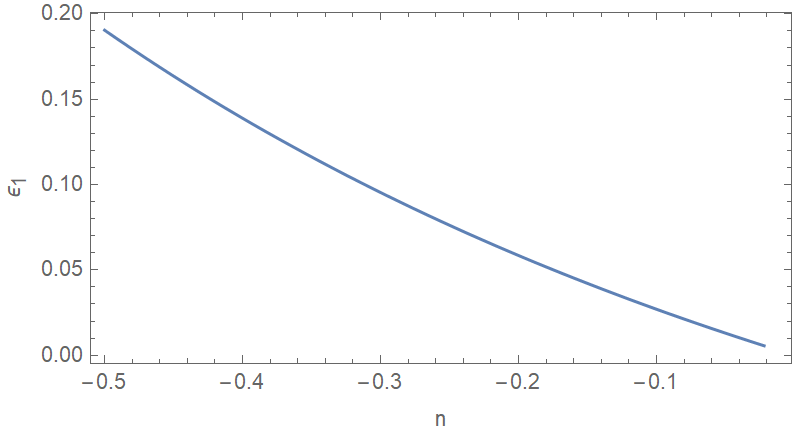}
\includegraphics[width=20pc]{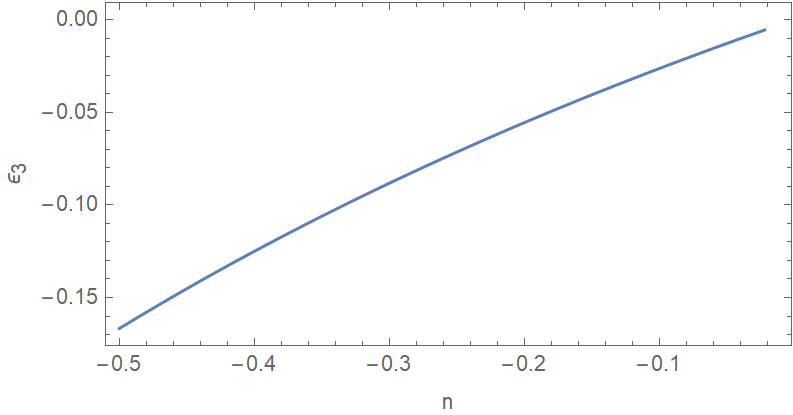}
\includegraphics[width=20pc]{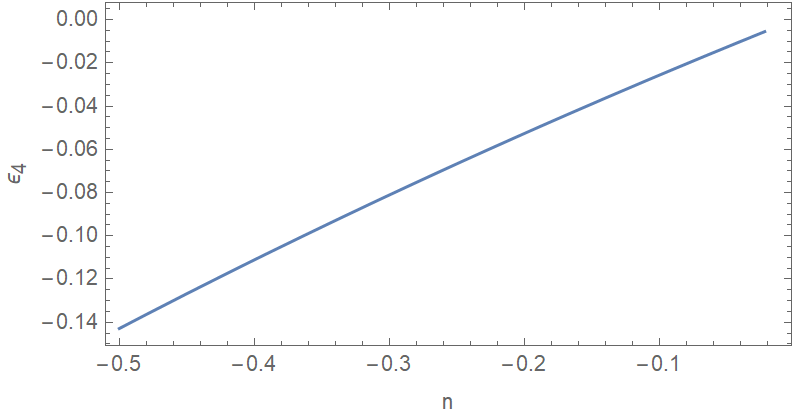}
\caption{The slow-roll indices $\epsilon_1$, $\epsilon_3$ and
$\epsilon_4$, for $n=[-0.038,-0.022]$. }\label{plot3}
\end{figure}

Now we can confront the power-law $F(R)$ gravity with the ACT data
\cite{ACT:2025fju,ACT:2025tim}, the Planck data
\cite{Planck:2018jri} and the updated Planck constraints on the
tensor-to-scalar ratio  \cite{BICEP:2021xfz}. One can easily see
that the theory is compatible with the Planck data for $n$ chosen
in the range $n=[-0.038,-0.03]$ and with the ACT data for $n$ in
the range $n=[-0.0282,-0.022]$. This can be seen in Fig.
\ref{plot1} where we plot the spectral index and the
tensor-to-scalar ratio parametric plot for $n$ in the range
$n=[-0.038,-0.022]$.
The confrontation of the model with the Planck and ACT data can
better be seen in Fig. \ref{plot2} where we present the
marginalized curves of the Planck 2018 data and the power-law
$F(R)$ gravity model confronted also with the ACT, and the updated
Planck constraints on the tensor-to-scalar ratio, for
$n=[-0.038,-0.022]$.
We can see in Fig. \ref{plot2}, that the power-law $F(R)$ gravity
model is well fitted within both the ACT and the updated Planck
data. To have a hands on grasp of the viability of the model, one
gets for $n=-0.025$, a spectral index $n_s=0.974365$ and a
tensor-to-scalar ratio $r=0.00194703$. Also let us demonstrate
that the slow-roll indices are smaller than unity during
inflation. Using Eqs. (\ref{epsilon1final2}),
(\ref{epsilon3final}) and (\ref{epsilon4finalnew}) in Fig.
\ref{plot3} we plot the values of the slow-roll indices for
$n=[-0.038,-0.022]$. As it can be seen, the values of the
slow-roll indices are indeed much smaller than unity.
Note that for the slow-roll index $\epsilon_3$ we used the its
leading order value $\epsilon_3\simeq -\frac{4+n}{4}\epsilon_1$.
In addition, the de Sitter stability criterion of Eq.
(\ref{criterion3}) is satisfied for the model at hand, because
$y=\frac{4+n}{4}$ at leading order for the model. Thus we
demonstrated that the power-law $F(R)$ gravity framework is a
viable inflationary phenomenological framework. Note that when the
Planck data are concerned, the Starobinsky deformations yield
almost indistinguishable results with the Starobinsky model. This
can better be seen in Fig. \ref{plot2}.

\subsection{The Case $n=0$: The Starobinsky Model}

What remains is to examine the parametrization we developed in the
previous sections for the case $n=0$. This case must yield the
correct behavior of the Starobinsky model, that is, a Planck
compatible quasi-de Sitter evolution. For $n=0$, the power-law
$F(R)$ gravity of Eq. (\ref{constnatxol}) becomes the $R^2$ model.
So for $n=0$, Eq. (\ref{epsilon1final2}) yields,
\begin{equation}\label{epsilon1starobinsky}
\epsilon_1=\frac{ c_3 R^{-1}}{c_1}\, ,
\end{equation}
which can be solved with respect to $H(t)$ and it yields,
\begin{equation}\label{quasidesitter}
H(t)=c_4-\frac{t}{12 c_1}\, ,
\end{equation}
where $c_4$ is an integration constant. Clearly the evolution
(\ref{quasidesitter}) is a quasi-de Sitter evolution, hence the
framework we developed yields the correct evolution for the
Starobinsky. Now let us show that it also confirm that the model
yields the correct phenomenology. The spectral index must be
evaluated at the first horizon crossing. Having $H(t)$ at hand,
one can evaluate explicitly the initial $t_i$ and final time $t_f$
instances of inflation. From $\epsilon_1=1$ we get $t_f=2 \left(6
c_1 c_2+\sqrt{3} \sqrt{c_1}\right)$ and also from the equation of
the $e$-foldings number,
\begin{equation}\label{efold}
N=\int_{t_i}^{t_f}H(t)\mathrm{d}t\, ,
\end{equation}
we easily get $t_i=2 \left(c_1 \sqrt{\frac{6
N}{c_1}+\frac{3}{c_1}}+6 c_1 c_2\right)$. Plugging the quasi-de
Sitter evolution (\ref{quasidesitter}) in Eq. (\ref{asxeto1}) and
(\ref{mainequation}) we get,
\begin{equation}\label{ns111}
n_s=1-\frac{48 c_1}{(t-12 c_1 c_2){}^2}\, ,
\end{equation}
and
\begin{equation}\label{r111}
r=\frac{6912 c_1{}^2}{(t-12 c_1 c_2){}^4}\, .
\end{equation}
So using Eqs. (\ref{ns111}) and (\ref{r111}), and plugging in the
initial horizon crossing time instance we obtain,
\begin{equation}\label{ns222}
n_s=\frac{2 N-3}{2 N+1}\, ,
\end{equation}
and
\begin{equation}\label{r222}
r=\frac{48}{(2 N+1)^2}\, .
\end{equation}
Upon expanding Eqs. (\ref{ns222}) and (\ref{r222}) at leading
order in the $e$-folding number $N$, we get,
\begin{equation}\label{ns333}
n_s\simeq 1-\frac{2}{N}+\frac{1}{N^2}-\frac{1}{2 N^3}\, ,
\end{equation}
and
\begin{equation}\label{r33}
r\simeq \frac{12}{N^2}-\frac{12}{N^3}\, .
\end{equation}
Both the scalar spectral index of Eq. (\ref{ns333}) and the
tensor-to-scalar ratio (\ref{r33}) describe the inflationary
phenomenology of the Starobinsky model at leading order. Thus we
demonstrated explicitly that our parametrization for the power-law
$F(R)$ gravity can reproduce the inflationary phenomenology of the
Starobinsky model, which is also a power-law $F(R)$ gravity.

In conclusion, the attributes of our parametrization for $F(R)$
gravity power-law inflation are:
\begin{itemize}\label{problems}
 \item The slow-roll approximation is not violated at first
 horizon crossing (the slow-roll indices are much smaller than
 unity).
 \item The analysis can produce the physics of the Starobinsky
 inflation, which is a power-law $F(R)$ gravity, in a natural way.
 \item Small deformations of the Starobinsky inflation are
 viable and respect the de Sitter stability criterion.
 \item Power-law $F(R)$ gravity inflation in our parametrization is
 viable with both the Planck and ACT data.
 \item In this parametrization, $\dot{\epsilon}_1\neq 0$, thus inflation is
 not eternal.
 \item In our parametrization, power-law $F(R)$ gravity and power-law
 evolution are disentangled. In fact, deformations of $R^2$
 inflation are quasi-de Sitter deformations.
\end{itemize}

\section{Conclusions}

In this work we aimed to re-address the power-law $F(R)$ gravity
inflation analysis, which in standard texts was perceived to be
non-viable, in the slow-roll approximation. We highlighted the
parametrization issues of the standard approach of power-law
$F(R)$ gravity inflation, which are the following: i) the
slow-roll approximation is violated during inflation$\cdot$ ii)
the analysis cannot reproduce the physics of the Starobinsky
inflation, which is actually a power-law $F(R)$ gravity with a
quasi-de Sitter evolution and not a power-law evolution$\cdot$
iii) small deformations of the Starobinsky inflation are not
viable and in fact, some of these do not even describe inflation,
which contradicts intuition$\cdot$ iv) In the standard
parametrization, $\dot{\epsilon}_1=0$, thus inflation is eternal
classically, and a power-law evolution. Now in our
parametrization, we highlighted what the problem is and we showed
how general $F(R)$ gravity inflation must be treated. Also we
applied our parametrization to power-law $F(R)$ gravity. In the
context of our parametrization, we found that the slow-roll
approximation is not violated at first horizon crossing since
slow-roll indices were found much smaller than unity. Also we
explicitly showed that our parametrization can produce the physics
of the Starobinsky inflation, which is actually a power-law $F(R)$
gravity, in a natural way. Also the evolution for the Starobinsky
model was found to be a quasi-de Sitter evolution, contrary to the
standard literature for power-law $F(R)$ gravity. Furthermore, we
found that small deformations of the Starobinsky inflation are
viable and in fact these deformations can be compatible with both
the Planck and ACT data. Also in our parametrization,
$\dot{\epsilon}_1\neq 0$, thus inflation is not eternal
classically. In fact, in our parametrization, power-law $F(R)$
gravity and power-law evolution are disentangled.

\section*{Acknowledgments}

This work was partially supported by the program Unidad de
Excelencia Maria de Maeztu CEX2020-001058-M, Spain (S.D.
Odintsov). This research has been funded by the Committee of
Science of the Ministry of Education and Science of the Republic
of Kazakhstan (Grant No. AP26194585) (S.D. Odintsov and V.K.
Oikonomou).

\end{document}